\documentclass[aps,prl,twocolumn,showpacs,floatfix,amsmath]{revtex4}

\usepackage{graphicx}
\begin{document}


\title{Intrinsic Tunneling Spectra of
Bi$_2$Sr$_2$CaCu$_2$O$_{8+\delta}$ near Optimal Doping}

\author{S. P. Zhao}
\author{X. B. Zhu}
\author{Y. F. Wei}
\affiliation{Beijing National Laboratory for Condensed Matter
Physics, Institute of Physics, Chinese Academy of Sciences,
Beijing 100080, China}


\begin{abstract}
We report tunneling spectra of near optimally doped
Bi$_2$Sr$_2$CaCu$_2$O$_{8+\delta}$ intrinsic Josephson junctions
with area of 0.09 $\mu$m$^2$, which avoid some fundamental
difficulties in the previous tunneling experiments and allow a
stable temperature-dependent measurement. A $d$-wave Eliashberg
analysis shows that the spectrum at 4.2 K can be well fitted by
considering electron couplings to a bosonic magnetic resonance
mode and a broad high-energy continuum. Above $T_c$, the spectra
show a clear pseudogap that persists up to 230 K, and a crossover
can be seen indicating two different pseudogap phases existing
above $T_c$. The intrinsic electron tunneling nature is discussed
in the analysis.
\end{abstract}


\maketitle

The paring mechanism and pseudogap phenomena are known as the two
key issues in the studies of high-$T_c$ superconductors. In
addition to the views based on resonating valence bond (RVB) state
\cite{lee06}, one possible mechanism that is of much interest is
boson-mediated pairing of the electrons. Recent angle-resolved
photoemission spectroscopy experiments have focused on a ``kink"
in single-particle dispersion and both phonon \cite{cuk04,zhou06}
and magnetic resonance \cite{john01,grom03} are considered as the
bosonic candidate. In the tunneling \cite{zasa} and optical
\cite{hwang04} experiments, a broad high-energy continuum, in
addition to the resonance mode, has also been considered.

Tunneling has traditionally been a useful tool in revealing the
material's superconducting properties. For the high-$T_c$
superconductors, mostly Bi$_2$Sr$_2$CaCu$_2$O$_{8+\delta}$
(Bi-2212), the scanning tunneling microscope (STM) \cite{renn98},
break junctions (BJs) \cite{zasa,miya98}, and intrinsic Josephson
junctions (IJJs) \cite{klei92,its,yurglaty,krasanag,zhu06,zhu07}
are used, yielding many important informations. In particular,
Zasadzinski {\it et al.} have correlated the tunneling dip
structure with the magnetic resonance mode \cite{zasa}. In Ref.
\onlinecite{renn98}, the pseudogap opening temperature
$T^{\star}$, rather than the supercondcting transition temperature
$T_c$, is suggested to be the mean-field critical temperature,
thus supporting precursor pairing as the origin of the pseudogap
phase.

IJJs, being intrinsic, have the advantage of avoiding the problems
like the surface chemical deterioration and unstable junction
structure, and can offer a convenient temperature-dependent
measurement. In the previous studies, however, the
micro-meter-sized IJJs are found to suffer from significant
self-heating, which severely distorts the tunneling spectra
\cite{krasanag,zhu06}.

\begin{figure}[b]
\centering
\begin{minipage}[c]{0.45\textwidth}
\scalebox{0.38}[0.38]{\includegraphics{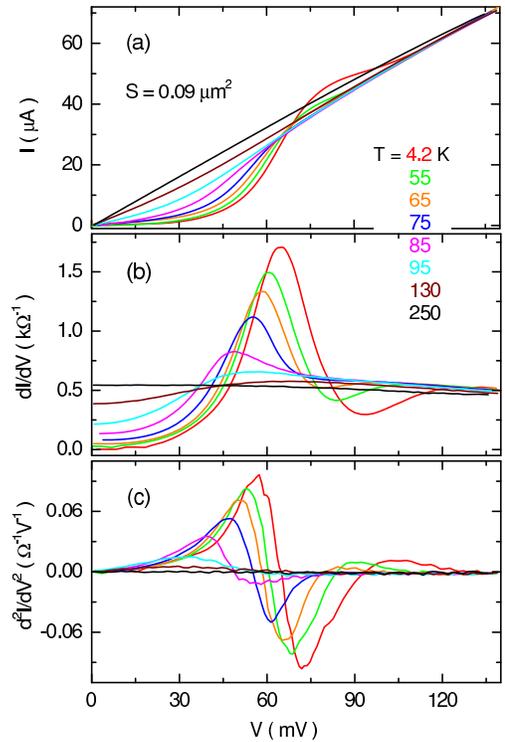}}
\end{minipage}
\caption{\label{fig:expiv} (color online). Temperature dependence
of the $I$-$V$, $dI/dV$, and $d^2I/dV^2$ curves of a Bi-2212 mesa
($T_c$ = 89 K) containing 10 IJJs. $V$ corresponds to the voltage
per IJJ.}
\end{figure}

Recently, we have demonstrated that self-heating in IJJs can be
reduced considerably when junction sizes decrease down to the
sub-micron level \cite{zhu06}. In this Letter, we report and
present a detailed analysis on the tunneling spectra of an IJJ
mesa with further-reduced area $S$ = 0.3$\times$0.3 $\mu$m$^2$ on
a near optimally doped Bi-2212 crystal \cite{rem1}, which are
shown in Fig.~\ref{fig:expiv}. Comparison with the previous
reports (including those from STM and BJ experiments) is presented
in Table~\ref{tab:table} in two important spectroscopic aspects:
the relation between the superconducting gap (SG) and pseudogap
(PG), and whether there is a clear dip at low temperatures. It can
be seen that self-heating can lead to considerably different
results in the IJJ experiments.

The spectra in Fig.~\ref{fig:expiv} show a well-defined, almost
temperature-independent normal-state resistance $R_N$, which allow
a straightforward tunneling analysis. Below we analyze the data,
along with a discussion on the tunneling nature in IJJs, of which
a consensus is still lacking. We will show that the spectrum at
4.2 K can be explained within the $d$-wave Eliashberg description
by considering a magnetic resonance mode plus a broad continuum
\cite{zasa}, with the latter having an upper cutoff at 300 meV
\cite{norm06}. Above $T_c$, there is a clear indication of the
existence of two pseudogap phases, which has not been reported so
far in the tunneling experiments for the (near) optimally doped
Bi-2212 samples.

\begin{table}[t]
\caption{\label{tab:table}Some differences and similarities of the
tunneling spectra for (near) optimally doped Bi-2212. Results for
STM, BJ, and IJJ are from Refs.~\onlinecite{renn98},
\onlinecite{miya98}, and \onlinecite{its}, respectively.}
\begin{ruledtabular}
\begin{tabular}{ccccc}
&STM &BJ &IJJ &This work \\
\hline
SG vs. PG\footnotemark[1]   &SG$\rightarrow$PG &no PG &SG/PG &SG$\rightarrow$PG \\
Clear dip?                  &yes               &yes   &no    &yes               \\
\end{tabular}
\end{ruledtabular}
\footnotetext[1]{SG$\rightarrow$PG and SG/PG denote a smooth
evolution from SG to PG at $T_c$ and that they coexist and evolve
independently, respectively.}
\end{table}

{\em The $d$-wave Eliashberg formalism}. --- We use a
boson-exchange spectral density in a separable form \cite{scha03}:
\begin{equation}
\alpha_{{\mathbf k},{\mathbf k^{\prime}}}^2F(\Omega)=
\alpha^2[~c_S+c_Dcos(2\theta) cos(2\theta^{\prime})~]{\mathcal F}
(\Omega)~,\nonumber
\end{equation}
\noindent where ${\mathbf k}$ or $\theta$ and ${\mathbf
k^{\prime}}$ or $\theta^{\prime}$ are associated with the initial
and final states in pairing interaction, $c_S$ and $c_D$ denote
the $s$- and $d$-wave components, and ${\mathcal F}(\Omega)$ is
the cut-off Lorentzians defined by \cite{scal69}
\begin{equation}
{\mathcal F}(\Omega)=\left\{\begin{array}{cc}
C\left[\frac{1}{(\Omega-\Omega_0)^2+\eta^2}-
\frac{1}{\Omega_1^2+\eta^2}\right],&|\Omega-\Omega_0|<\Omega_1~,\\
\\
0~,~~~~~~~~~~~~~~~~~~~~~~~~~~~&|\Omega-\Omega_0|>\Omega_1~,
\end{array}
\right.\nonumber
\end{equation}
\noindent in which $C$ normalizes ${\mathcal F}(\Omega)$ to unity.
The Eliashberg equations in the $d$-wave form at $T$ = 0 can be
written as:
\begin{equation}
\Delta(\omega)=\frac{c_D}{Z(\omega)}
\int\limits_{0}^{\infty}d\omega^{\prime}
\int\limits_{0}^{2\pi}\frac{d\theta^{\prime}}{2\pi}
Re\left[\frac{\Delta(\omega^{\prime})cos^22\theta^{\prime}K_{+}(\omega,\omega^{\prime})}
{\sqrt{\omega^{\prime
2}-\Delta^2(\omega^{\prime})cos^22\theta^{\prime}}}\right],\nonumber
\end{equation}
\begin{equation}
Z(\omega)=1-\frac{c_S}{\omega}
\int\limits_{0}^{\infty}d\omega^{\prime}
\int\limits_{0}^{2\pi}\frac{d\theta^{\prime}}{2\pi}
Re\left[\frac{\omega^{\prime}K_{-}(\omega,\omega^{\prime})}
{\sqrt{\omega^{\prime
2}-\Delta^2(\omega^{\prime})cos^22\theta^{\prime}}}\right]\nonumber
\end{equation}
\noindent in which $\Delta$ and $Z$ are the gap and
renormalization functions, and $K_{\pm}(\omega,\omega^{\prime})$
is given by
\begin{eqnarray}
K_{\pm}(\omega,\omega^{\prime})= \int\limits_{0}^{\infty}d\Omega~
\alpha^2{\mathcal F}(\Omega) ~(~\frac{1}
{\omega^{\prime}+\omega+\Omega-i\delta} ~~~~~~~~ \nonumber \\
~\pm \frac{1}{\omega^{\prime}-\omega+\Omega-i\delta}~)
~.~\nonumber
\end{eqnarray}
The above equations can be solved if the parameters $c_S$, $c_D$,
and $\alpha^2{\mathcal F}(\Omega)$ are known. Once
$\Delta(\omega)$ and $Z(\omega)$ are obtained, the diagonal
self-energy of the system can be found from Re$\Sigma$ = $\omega
[1-$Re$Z]$ and Im$\Sigma$ = $-\omega$Im$Z$, and the density of
states (DOS) is given by
\begin{equation}
N(\omega,\theta)=Re[\omega/
\sqrt{\omega^2-\Delta^2(\omega)cos^2(2\theta)}]~.\nonumber
\end{equation}

\begin{figure}[b]
\centering
\begin{minipage}[c]{0.48\textwidth}
\scalebox{0.45}[0.45]{\includegraphics{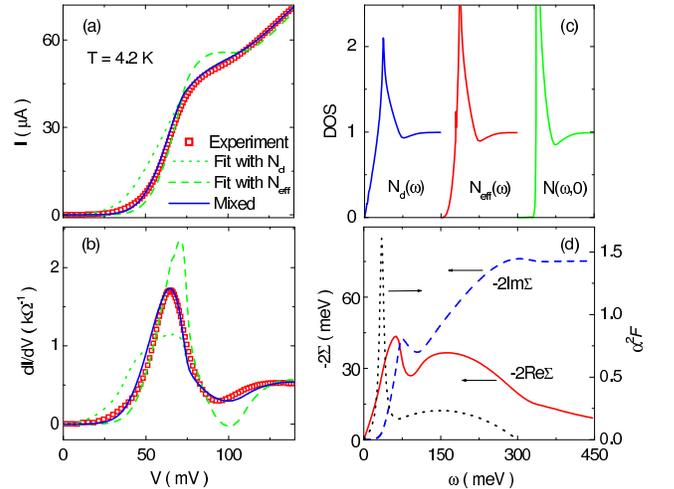}}
\end{minipage}
\caption{\label{fig:iv4k} (color online). Fit of the experimental
(a) $I$-$V$ and (b) $dI/dV$ curves at 4.2 K using the $d$-wave
Eliashberg theory. The calculated DOS and self-energy $\Sigma$ are
shown in (c) and (d). Note the horizontal shifts of 150 and 300
meV for $N_{eff}$ and $N(\omega,0)$ in (c) for clarity.
$\alpha^2{\mathcal F}$ used in the calculation is also shown in
(d). See text for detailed fitting parameters.}
\end{figure}

{\em Discussion on the tunneling nature in IJJs}. --- A usual way
to fit the tunneling spectra is to build an angle-averaged DOS:
$N_d(\omega)$ = $(1/2\pi)$
$\int\limits_{0}^{2\pi}N(\omega,\theta)d\theta$, from which the
$I(V)$ curve can be calculated straightforwardly considering an
incoherent tunneling process \cite{zasa,esch00,rem2}. The basic
features from such calculation at 4.2 K using a set of parameters
described below can be seen in Fig. \ref{fig:iv4k}(a) and (b)
(dotted lines). While the lines follow the experimental data
closely above the gap voltage, they deviate considerably below it.
From our calculation, we find that a better fit below the gap
voltage can be obtained by considering coherent electron tunneling
with a directional matrix element of $t$ $\sim$ $cos^2(2\theta)$,
but this leads to a worse agreement near and above the gap voltage
\cite{esch00}.

Coherent electron tunneling has been discussed in a number of
publications \cite{esch00,xiang96,marel99} and used to interpret
the $c$-axis transport data in IJJs \cite{laty99,zhao05}. The
present results may imply that there is a crossover from coherent
to incoherent tunneling as the bias voltage increases. Difficulty
arises, however, that when a $t$ $\sim$ $sin^2(4\theta)$ is used,
which is theoretically more strict for the Bi-2212 material in
which the Cu atoms in CuO$_2$ planes across the BiO and SrO
barrier layers do not lie collinearly \cite{marel99}, the computed
results give a worse fit for voltages both below and above the gap
edge. To avoid the conceptual difficulty, we assume that intrinsic
electron tunneling in IJJs is basically incoherent for voltages
away from zero. At low bias, tunneling can be described by an
effective DOS: $N_{eff}(\omega)$ = $(1/\pi)$
$\int\limits_{0}^{2\pi}N(\omega,\theta) cos^2(2\theta)d\theta$,
which produces similar $I$-$V$ and $dI/dV$ characteristics (dashed
lines in Fig. \ref{fig:iv4k}(a) and (b)) as coherent tunneling
plus $t$ $\sim$ $cos^2(2\theta)$ is considered. Physically, the
$cos^2(2\theta)$ factor in $N_{eff}$ can result from the
wave-function overlap (having $d_{x^2-y^2}$ symmetry) between the
O-2p and Cu-4s orbitals while the latter plays an important role
in assisting a hole hopping to and out of the O-2p orbital
\cite{xiang96}. At high bias, tunneling is described by
$N_d(\omega)$.

Switch from $N_{eff}$ to $N_d$ implies a lesser role the Cu-4s
orbital plays in assisting hopping as bias increases and more
high-energy quasiparticles with reduced lifetimes are involved in
tunneling. Phenomenologically, the tunneling current can be
written as $I$ = $PI_d$ + $(1-P)I_{eff}$, where $I_d$ and
$I_{eff}$ represent those obtained using $N_d$ and $N_{eff}$, and
$P$ is a voltage-dependent probability distribution. In our
fitting, an exponential function $P(V)$ =
($e^{V/V_1}-1$)/($e^{V_2/V_1}-1$) with $0\le V\le V_2$ is used, in
which $V_1$ characterizes the voltage scale for $I_{eff}$ to $I_d$
crossover and $V_2$ sets a point above which $I$ = $I_d$. The
solid lines in Fig. \ref{fig:iv4k}(a) and (b) are obtained from
the dotted and dashed results with $V_1$ and $V_2$ equal to 20 and
75 mV, respectively. The fit appears satisfactory.

{\em Tunneling spectrum at 4.2 K}. --- In the Eliashberg analysis
of the 4.2-K data in Fig.~\ref{fig:iv4k}, we follow Zasadzinski
{\it et al.} and use two Lorentzians for $\alpha^2{\mathcal F}$ to
model the magnetic resonance peak (around 35$\sim$40 meV) and the
high-energy continuum \cite{zasa}. The fitting parameters are:
$c_S$ = 0.18, $c_D$ = 1; $\alpha^2_1$ = 20, $\Omega_{01}$ =
$\Omega_{11}$ = 35 meV, $\eta_1$ = 5 meV; $\alpha^2_2$ = 46.5,
$\Omega_{02}$ = $\Omega_{12}$ = 150 meV, $\eta_2$ = 2000 meV;
$R_N$ = 1.92 k$\Omega$. The parameters are adjusted so as to have
a fair fit of the voltage positions of the $dI/dV$ peak and dip
and also the depth of the dip, which are therefore judged
primarily by $I_d$. The resulting antinodal DOS $N(\omega,0)$,
together with $N_d$ and $N_{eff}$, is shown in
Fig.~\ref{fig:iv4k}(c), and $\alpha^2{\mathcal F}$ is depicted in
(d).

$\alpha^2{\mathcal F}$ is found to change the shape of the $dI/dV$
curve in the following way. While the resonance peak creates the
dip, the high-energy continuum tends to reduce its depth. In
addition, as the continuum extends to higher energy, the width of
the dip decreases. The latter change is more obvious when the
upper cutoff of the continuum reduces from 150 to 80 meV, where
the resulting dip width in both the $dI/dV$ and DOS curves becomes
too wide to fit the experimental data. This means that a continuum
extending above 150 meV must be considered.

We have chosen an upper cutoff of 300 meV for the continuum
following a recent analysis of the optical data \cite{norm06}. As
can be seen in Fig.~\ref{fig:iv4k}(d), this results in a
considerable high-energy contribution to $-2$Re$\Sigma$, which may
need further experimental testing. Using a smaller cutoff of
$\sim$ 160 meV \cite{zasa} will reduce such contribution and
slightly increase the dip width. The overall size of
$-2$Re$\Sigma$, on the other hand, is mostly affected by $c_S$.
Reducing $c_S$ will reduce $-2$Re$\Sigma$ and make the peak-dip
separation smaller.

Compared with the BJ results \cite{zasa}, the present spectra show
a shallower dip, which leads to a larger contribution of the
continuum in $\alpha^2{\mathcal F}$ relative to the resonance
mode. Presently, it seems unclear as to which part in
$\alpha^2{\mathcal F}$ should be more essential for the pairing
\cite{zasa,hwang04}.


\begin{figure}[t]
\centering
\begin{minipage}[c]{0.45\textwidth}
\scalebox{0.4}[0.4]{\includegraphics{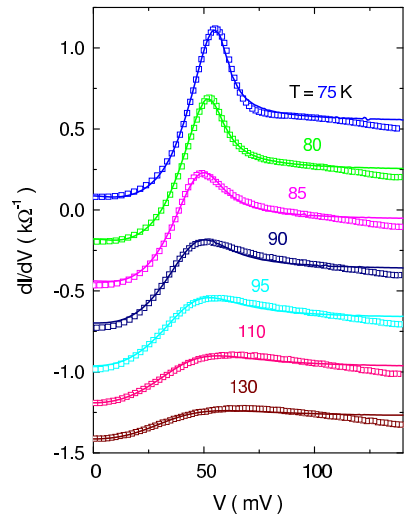}}
\end{minipage}
\caption{\label{fig:didvtc} (color online). Experimental $dI/dV$
curves near and above $T_c$ (symbols). Lines are fits using
$N_{eff}$, in which $N(\omega,\theta)=Re[(\omega -i\Gamma)/
\sqrt{(\omega -i\Gamma)^2-\Delta^2cos^2(2\theta)}]$, with $\Delta$
and $\Gamma$ values plotted in Fig.~\ref{fig:gapvst}. Curves are
shifted downwards successively by 0.3 $k\Omega^{-1}$ for clarity
except for that at 75 K.}
\end{figure}

\begin{figure}[t]
\centering
\begin{minipage}[c]{0.4\textwidth}
\scalebox{0.35}[0.35]{\includegraphics{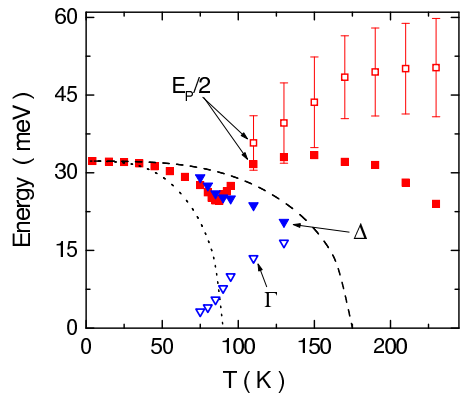}}
\end{minipage}
\caption{\label{fig:gapvst} (color online). Peak energy $E_P$
divided by two from experimental $dI/dV$ curves (solid squares).
Open squares are the results from the $dI/dV$ curves normalized to
the 250-K curve. Solid and open triangles are the $\Delta$ and
$\Gamma$ values used for fitting in Fig.~\ref{fig:didvtc}. The
dashed line is $\Delta (T)$ from the BCS $d$-wave description with
$2\Delta(0)/kT_c$ = 4.3. The dotted line is the corresponding
result with temperature normalized to $T_c$.}
\end{figure}

{\em Spectra near and above $T_c$}. --- As temperature increases,
the dip gradually disappears at $\sim$ 75 K (see
Fig.~\ref{fig:expiv}(b) and (c)). In this case, it is possible to
fit the data using $N_{eff}$ in which $N(\omega,\theta)$ has a
Dynes' form $Re[(\omega -i\Gamma)/ \sqrt{(\omega
-i\Gamma)^2-\Delta^2cos^2(2\theta)}]$ where $\Delta$ and $\Gamma$
are the energy-independent parameters \cite{dynes78}. The fitting
can be satisfactory using $N_{eff}$ in the entire voltage range
since difference of using $N_{eff}$ and $N_d$ above the gap
voltage is smaller at higher temperatures and in the absence of
the dip. Comparison between experiment and theory can be seen in
Fig.~\ref{fig:didvtc}, and the $\Delta$ and $\Gamma$ parameters
are plotted in Fig.~\ref{fig:gapvst}. It is interesting to note
the linear temperature dependence of the scattering rate $\Gamma$,
a signature predicted in the marginal Fermi-liquid theory
\cite{varma89}.

It should be pointed out that using $\Delta$ to describe the data
above $T_c$ implies that we are taking the pre-pairing view and
$\Delta$ is understood as the thermodynamic-averaged value
\cite{tink96}. The pre-pairing idea has been suggested previously
in the STM experiments to account for the pseudogap phase
\cite{renn98}. From our experiments, the $dI/dV$ spectra show a
gapped structure for temperatures as high as 230 K. To look at
this in more detail, we plot in Fig.~\ref{fig:gapvst} the
temperature-dependence of half the peak energy in the $dI/dV$
curves ($E_P$/2: solid squares). It can be seen that $E_P$/2
decreases as the temperature increases up to $T_c$. It then
increases, reaches a maximum at $\sim$ 140 K, and starts
decreasing again after that. The open squares in the figure are
the results taken from the $dI/dV$ curves that are normalized to
the 250-K curve. In this case, as temperature decreases from 230
K, $E_P$/2 is almost flat until $\sim$ 170 K, below which it
decreases at a relatively fast speed.

These results point to a change within the pseudogap phase, which
can be characterized by two temperatures $T_1^{\star}$ = 230 K and
$T_2^{\star}$ = 140$\sim$170 K. Considering our near optimally
doped sample, this is consistent in the phase diagram with the
recent Nernst experiment \cite{yayu05}. From the RVB scenario
\cite{lee06}, the results can be viewed as the opening of a spin
gap below $T_1^{\star}$. Below $T_2^{\star}$, holons start to
Bose-condense and the Nernst state results. A global
superconducting state is finally developed below $T_c$ through the
Berezinskii-Kosterlitz-Thouless transition. In the view of
boson-mediated pairing, the change around $T_2^{\star}$ should
correspond to the point below which the superconducting order
starts to form out of (with $\Gamma$ comparable to $\Delta$ at
initial stage), and compete with the spin-gap state.

{\em Summary}. --- Tunneling spectra of near optimally doped
Bi-2212 IJJs in the temperature range of 4.2 to 250 K are
presented. We have analyzed intrinsic tunneling process and used
$d$-wave Eliashberg theory and Dynes' DOS to fit the tunneling
data. The fit between theory and experiment is satisfactory. We
have shown that the electronic coupling to a magnetic resonance
mode plus a broad continuum can well explain the spectrum at 4.2
K, thus providing an evidence for the electronic-originated
boson-mediated pairing. However, a doubt remains as to which part
in the boson spectrum is more essential for superconductivity. We
have also found that there exist two pseudogap phases above $T_c$.

We acknowledge many valuable discussions with T. Xiang, N. L.
Wang, and Q. H. Wang. This work was supported by the Ministry of
Science and Technology of China (2006CB601007), the Knowledge
Innovation Project of the Chinese Academy of Sciences, and the
National Natural Science Foundation of China (10604064).


\begin{thebibliography}{}
\bibitem{rem1}The mesa contains 10 IJJs, with bulk $T_c$ of 89 K and a
$T_c^{\prime}$ of 71 K for the surface layer. Maximum temperature
rises are estimated to be only 2.5 and 5.5 K at the biases
corresponding to the conductance peak and dip, respectively. Part
of the data in magnetic field is presented in Ref.
\onlinecite{zhu07}.
\bibitem{rem2} In the incoherent case,
the tunneling current $I(V)$ =
$(eR_N)^{-1}\int\limits_{-\infty}^{\infty}n(\omega)n(\omega^{\prime})
\left[f(\omega)-f(\omega^{\prime})\right]d\omega$, where
$\omega^{\prime}$ = $\omega$ + $eV$ and the tunneling DOS
$n(\omega)$ depends only on energy. Both $N_d$ and $N_{eff}$ are
considered for $n(\omega)$ in our discussion.
\end{thebibliography}
\end{document}